# Dynamical Disequilibrium, Transformation, and the Evolution and Development of Sustainable Worldviews

**Liane Gabora and Maegan Merrifield**
University of British Columbia

This chapter begins by outlining a promising, new theoretical framework for the process by which human culture evolves inspired by the views of complexity theorists on the problem of how life began. Elements of culture, like species, evolve over time; that is, they exhibit cumulative change that is adaptive in nature. By studying how biological evolution got started, we gain insight into not just the specifics of biological evolution, but also general insights into the initiation of any evolutionary process that may be applicable to culture. We then explore the implications of this new framework for culture on the transformative processes of individuals. Specifically, we will address what this emerging perspective on cultural evolution implies for to go about attaining a *sustainable worldview*; that is, a web of habits, understandings, and ways of approaching situations that is conducive to the development of a sustainable world.

**How Does an Evolutionary Process Get Started?**

In attempting to gain insight into the origins of transformative processes in individuals, it is instructive to look at the transformative processes by which the earliest forms of life evolved. Research into how life began is stymied by the improbability of a spontaneously generated structure that replicates itself using a self-assembly code such as the genetic code. Something so complex as this is unlikely to emerge out of the blue! Hoyle (1981) infamously compared it to the probability that a tornado blowing through a junkyard would assemble a Boeing 747.

This has led to widespread support for different versions of the proposal that the earliest self-replicating structures were autocatalytic sets of molecules (Bollobas, 2001; Bollobas & Rasmussen, 1989; Dyson, 1982, 1985; Gabora, 2006; Kauffman, 1986, 1993; Morowitz, 1992; Segré, Ben-Eli, & Lancet, 2000; Segré, et al, 2001a, b; Wäechtershäeuser, 1992; Weber, 1998, 2000; Williams & Frausto da Silva, 1999, 2002; Vetsigian *et al*., 2006). A set of molecules is *autocatalytic* if every molecule in the set can be regenerated through chemical reactions occurring amongst *other* molecules in the set. (The term 'autocatalytic' comes from the fact that the molecules speed up or *catalyze* the reactions by which other molecules are formed.) Reactions between molecules generate new, different molecules. As the number of different molecules increases, the number of reactions by which they can interconvert increases even faster (Cohen, 1988; Erdös & Rényi, 1960). Thus some subset of them reaches a critical threshold where there is a reaction pathway to the formation of every molecule in the set.

At this point, the parts can reconstitute the whole in a piecemeal manner, through bottom-up interactions rather than top-down interpreting of a genetic code (Kauffman, 1993; for summary see Gabora, 2008). The hydrophilic (water-loving) molecules orient toward the periphery, forming a spherical vesicle that encloses the more hydrophobic (water-avoiding) molecules. This kind of spherical vesicle made up of collectively self-replicating parts is sometimes referred to as a *protocell*. It is prone to fission or budding, wherein part of it pinches off, and it divides in two. So long as there is at least one copy of each polymer in each of the two resulting vesicles, they can self-replicate, and continue to do so indefinitely, or until their structure changes drastically enough that self-replication capacity breaks down, and by that point there will exist other self-replicating sister-lineages. The process is sloppier and more haphazard than the self-replication that occurs in modern day organisms. Some raise the concern that, at least with respect to some versions of this theory, replication occurs with such low fidelity that evolvability breaks down (Vasas, Szathmary, & Santos, 2009). Nevertheless, there is broad consensus that such a structure remains sufficiently intact over generations for stable evolution (Schuster, 2010). A key thing to note here is that with this kind of self-replication there is nothing to prohibit the inheritance of acquired characteristics. A change to any one part of the structure persists after fission occurs, and this may cause other changes that have a significant effect further downstream.

Evolution of these early life forms occurs through horizontal exchange (i.e. not restricted to vertical transmission from parent to offspring) of "innovation-sharing protocols" (Vetsigian *et al.,* 2006). It was not until the genetic code came into existence—and the process in which self-assembly instructions are copied (meiosis) became distinct from developmental processes—that acquired characteristics could no longer be passed on to the next generation (Gabora, 2006). The work of Woese and his colleagues indicates that early life underwent a transition from this fundamentally cooperative process horizontal evolution through communal exchange, to a fundamentally competitive process of vertical evolution by way of the genetic code. This transition is referred to as the *Darwinian threshold* (Woese, 2002) or *Darwinian transition* (Vetsigian et al., 2006) because it marks the onset of what we think of as conventional Darwinian evolution through natural selection. Kalin Vetsigian (pers. comm.) estimates that the period between when life first arose and the time of the Darwinian threshold spanned several hundred million years.

Thus we have two kinds of self-replication (Gabora, 2004). *Coded self-replication,* such as is seen in present-day organisms, uses self-assembly instructions as proposed by von Neumann. This ensures they replicate with high fidelity, and acquired characteristics are not inherited. *Uncoded self-replication,* such as is seen in protocells, involves autocatalysis. This is a low fidelity means of replication, and there is nothing to prohibit inheritance of acquired characteristics. Note that it is sometimes said that because acquired traits are inherited in culture, culture cannot be described in evolutionary terms. It is ironic that scientists working on the problem of how life began have shown that this is also the case with respect to the earliest stage of biological life itself. These origin of life scientists have however resolved the problem of how culture could evolve despite inheritance of acquired characteristics by showing that it is possible for an evolutionary process to get underway and for adaptive, cumulative change to take place *without* natural selection.

In sum, there is increasing evidence that the very earliest life forms were self-organized metabolisms that self-replicated rather haphazardly through duplication of their catalytic components. They evolved through a non-Darwinian (Lamarckian) process involving piecemeal transformation and communal exchange, as opposed to the competitive exclusion or 'survival of the fittest' that characterizes natural selection.

**How does Human Culture Evolve?**

Over time our widgets have become more complex and our art has become more varied, leading many to suggest that human culture evolves. However it has not been definitively established in what sense culture constitutes an evolutionary process. To do so is the overarching goal of the various strands of my research. I aim to bring forward a theoretical framework for cultural evolution that is as sound as our theoretical framework for biological evolution, and apply it to the tasks of reconstructing our past, exploring possible futures, and furthering human wellbeing.

It is sometimes assumed that the self-replicating units of cultural evolution are artifacts (tools, fashions, etc.) or the ideas or 'memes' that give rise to them, and that they evolve, like modern-day organisms do, through natural selection. But as pointed out earlier, von Neumann showed that the key feature of self-replication in biology is not mere self-copying, but self-copying using a code (such as the genetic code) that functions both as a self-description (passively transmitted to offspring during the process of reproduction) and as self-assembly instructions (actively interpreted to build the organism during the process of development). It is because of this division of labour that in biological evolution, acquired traits are not inherited (e.g., you do not inherit your mother's tattoo). However, there is nothing in cultural evolution that functions both as self-description and self-assembly instructions. That is why acquired characteristics are transmitted. For example, if you are told a joke and you think of a funnier version of the joke it is probably your version that you tell to someone else; the new trait acquired by the joke is transmitted. This is one reason why natural selection does not provide an adequate explanatory model of cultural evolution.

An alternative to the notion that culture is Darwinian is the proposal that the transformative, evolutionary, adaptive processes of culture are structurally similar to those of the earliest life forms (Gabora, 2000, 2004, 2008, in press; Gabora & Aerts, 2009). It is proposed that in human culture what evolves is not discrete ideas or artifacts but minds, or more specifically, peoples' internal models of the world, including knowledge and how it has been made sense of, as well as ideas, hopes, attitudes, beliefs, values, predispositions, and habitual patterns of thought and behavior. This internal model of the world is referred to as a *worldview*. Worldviews evolve in the same haphazard sense as these earliest life forms, not through competition for survival of the *fittest* as modern-day life, but through transformation of *all*. In other words, the assemblage of human worldviews changes over time not because some replicate at the expense of others, as in natural selection, but because of ongoing mutual interaction and modification.

A human worldview has the following properties. It is self-organizing in the sense that one constantly uses newly acquired information to update one's general understanding of people and things and how they are related. It is self-mending in the sense that just as a body spontaneously heals itself when wounded, if something unexpected happens one can't help but try to figure out why, i.e. to revise one's worldview to accommodate this unexpected event. A worldview is also autopoietic, i.e. the whole emerges through interactions amongst the parts. As a young child thinks through how the different bits of information it learns are related to one another it comes to have an integrated understanding of the world it lives in and his or her relationship to that world.

We saw that an important component of an evolutionary process is self-replication—basically, making a copy of oneself. An adult shares ideas, stories, and attitudes with children (and other adults), influencing little by little the formation of other worldviews. Of necessity, a worldview acquires and expresses cultural information in the form of discrete units (e.g. gestures or artifacts), but the processing of it (i.e., the process by which it acquires characteristics) reflects its own multifaceted web of knowledge, experience, needs, and perspectives. Different situations expose different facets of a worldview (much like cutting a fruit at different angles exposes different parts of its interior). Elements of culture (such as rituals, customs, beliefs, and artifacts) reflect the current evolutionary states of the worldviews that generate them. People influence the contexts by which the potential of their worldviews is realized, either externally, by influencing their environments, or internally, by generating their own contexts (e.g. fantasy, rumination, counterfactual thinking).

**Implications of the Theory for Personal Transformation**

This theory of cultural evolution has significant implications for the individual's process of therapeutic self-transformation, which it is argued is a creative process that arises due to the *self-organizing, self-mending* nature of a worldview. In other words, the ruminative, associative thought processes that one engages in when in a state of emotional upheaval or internal dynamical disequilibrium reflects the natural tendency of a worldview to seek integration or consistency amongst both its pre-existing and newly-added components, including ideas, attitudes, or bits of knowledge (Gabora, under revision). Each idea or interpretation of a situation that the individual comes up with is a different expression of the same underlying core network of understandings, beliefs, and attitudes. A worldview has a characteristic structure, and the individual's behavior reflects the (to some extent) unique architecture of this web of understandings.

The individual's behavior and other outputs are like footprints in the snow, suggesting the 'shape' and dynamics of the underlying worldview that gave rise to them but which we never see directly. The individual's behavioral outputs are thus integrally related to one another and potentially pave the way for one another (Gabora, 2010; Gabora, O'Connor, and Ranjan, in press). They also pave the way for the transformative processes of others who interact with that individual and/or his or her creative outputs, and thereby the transformative process of each individual contributes to the global process of the human cultural evolution.

It is proposed that what drives this is a sense of fragmentation or dynamical disequilibrium. This may be externally generated, i.e. it may arise due to life problems, knowledge or ideas that do not cohere, unexpected situations, injustices, and so forth. Alternatively, the dynamical disequilibrium may be internally generated, *i.e*., it may arise through intrinsically motivated play or exploration, perhaps driven by curiosity or restlessness. Because a human worldview is self-organizing, people influence the contexts by which the potential of their worldviews is realized, and thereby respond to and 'mend' the state of disequilibrium. Though imagination, fantasy, rumination, and 'what if' type thinking, further play and exploration, or trying things out, a worldview is exposed to contexts that may transform to a more integrated state. The individual thereby regains dynamical equilibrium (absence of major life problems, knowledge and ideas more or less cohere, unexpected situations explained, injustices put right, and so forth).

Thus, the therapeutic self-transformation arises as a result of the propensity of an internal model of the world to respond to dynamical disequilibrium with ruminative, associative processes aimed at self-reorganization. The transformative processes may take place in an informal setting, through reflection, perhaps aided by discussion, or an experience of nature. It may also occur in a formal setting such as through art therapy.

The American Art Therapy Association (2008) defines art therapy as a clinical practice "being dedicated to the belief that making art is healing and life enhancing." It is based on the idea that there is an inherent healing power in the creative process and that art making provides access to things that are difficult or impossible to verbalize. Art therapists believe that there are two processes going on during the act of creating : 1) the work produced through art making within the clinical sessions, and 2) its accompanying emotional catharsis or connection. The creative work can provide a springboard for discussion between the client and therapist, and it can also be analyzed for symbolic meaning and used in assessment (Malchiodi, 2007).

Art therapy is comprehensive in its objectives. Some goals of art therapists are to help people find healthier ways of dealing with their personal internal processes, other people, their community, and their environment (Dunn-Snow & Smellie, 2011). Therapy can help build on skills activated in the past (Riley, 1999). Clinical experience indicates that creating art is a way to manage intense feelings as a result of emotional work (Moon, 1999). Experimental studies of emotion and creativity indicate that higher levels of creativity are correlated with positive affect (Hennessey & Amabile, 2010). Clinical practitioners of art therapy note that imagery and creative engagement can deepen communication between the client and therapist (Moon, 2009). A primary goal of therapy is to equip clients with skills that transfer to other settings, including communication skills. It can also offer a way for the clients to achieve insights into situations and behaviors that could not be obtained by thinking and speaking in a logical manner (Moon, 2009). Art can thereby open doors to healing and create a free and safe space to express emotional experience.

The honing theory of creativity can also be applied to art therapy. Honing theory places equal emphasis on the externally visible creative outcome and the internal cognitive restructuring brought about by the creative process. Art therapy has the potential to address this self-mending process. Honing involves drawing associations, reorganizing understandings, and working through emotions. The more experiences one has or tensions one has to face, the more ingredients one has to draw upon in this process. Artistic expression can be seen as the manifestation of deeper hidden structures, and as a means of accessing personal history that would not be available through verbalization (Karkou & Sanderson, 2006). The process of creating can result in new associations and understandings that pave the way toward more constructive attitudes, and thus enhanced mental health. This can lead to better coping. Through the use of art in a therapeutic context, the individual can restructure negative thoughts and experiences into positive ones. Art therapy (as well as related therapies such as music therapy and dance therapy) can thus provide a structured way for people to respond to and challenge beliefs and attitudes, and develop an integrated worldview.

**Education and the Compartmentalization of Knowledge**

What are the implications of this new view of personal transformation founded on a scientific perspective on the manner in which culture evolves? One such implication has to do with how we teach children about the world they live in. The view of reality that is presented to children both aids and constrains how that child's worldview develops. We present the world to children in a compartmentalized fashion. Spontaneous deviations from the daily school routine are not tolerated because curricula would not be covered, and scores on standardized tests would fall. Even the weekends offer little possibility for spontaneously playing with, exploring, and discovering for oneself an understanding of the world that strays from the compartmentalized version of reality we are handed down. Moreover, if children do discover something new for themselves, they are often discouraged from exploring it further. As a child, one of the authors of this chapter (for simplicity, hereafter referred to as "I") developed a "framework for all things". It included: things that repeatedly go back and forth unchanged (such as a pendulum or metronome), things that repeatedly go back and forth but change (such as night and day, or a spiral), things that just go up and down (such as a ball thrown into the air), things that go up and down but with a change (such as the feelings you have about a story during and after reading it). I was made to feel like what I had arrived at was not the correct way of classifying elements of life. I had gotten it all wrong. The "correct" way was in terms of the subjects you learn at school, and the jobs adults have: math, art, biology, social studies, and so forth.

It wasn't until much later that I started to question to what extent these conventional categories objectively capture something *real* about the world (and to wonder why no one else seemed to find this important.) Most of my friends at the time were computer scientists working in the entertainment industry developing equations and algorithms to render animated creatures in life-like, artistically appealing, socially acceptable ways. So much for the view that math, art, biology, and social studies are distinct and separate domains! Life doesn't *come* compartmentalized; *any* way one goes about carving up reality involves a degree of arbitrariness.

**A Sustainable Worldview is an Integrated Worldview**

It is useful, sometimes essential, to categorize and compartmentalize in order to accomplish particular tasks. But if we take any sort of compartmentalization seriously, that is, if we assume it to be a faithful representation of what really exists, it distorts how we see the world, which in turn distorts how we act in it. Behind a sustainable world lies a sustainable worldview. By *sustainable worldview*, I mean a way of seeing the world and being in the world that incorporates how different systems are interconnected and mutually affect one another, an internal model of the world that is *ecological* in character. Sustainability requires not just understanding and solutions; it requires the laying down of new habitual patterns of thinking and acting that foster sustainable outcomes, and that over time become second nature (Clayton & Brook, 2005; Gifford, 2006; Koger & Winter, 2010; Kurz, 2002 Nickerson, 2003; Vlek & Steg, 2007; Winter & Koger, 2004). The compartmentalization of knowledge may lead to an artificially compartmentalized view of the world, which may in turn interfere with our capacity to think about the multifaceted downstream consequences of our actions.

If knowledge is *presented* in compartmentalized chunks, then our youth may end up with a compartmentalized understanding of the world. If knowledge were presented more holistically, a more integrated kind of understanding may be possible. Exposure to a wide range of creative opportunities such as dance, painting, pottery, or creative writing allows for divergent thinking and novel connections to be made, thus encouraging integration of many experiences into a cohesive view. It may be that our potential for a deeply ecological worldview in a modern context is just beginning to be exploited, through ventures such as the Learning through the Arts program, in which students, for example, learn mathematics through dance, or learn about food chains through the creation of visual art. This program originated in Canada but is now being adopted abroad. Such activities can contribute to the therapeutic self-transformation of the individual by providing new narrative structures forging a more reliable, resilient, and empathic connection between the various parts of oneself, and the world in which one lives.